\documentclass[twocolumn]{autart}

\usepackage[round]{natbib}
\usepackage{amsmath,bm}
\usepackage{color}
\usepackage{multirow}
\usepackage{mathrsfs}
\usepackage{tcolorbox}
\usepackage{subfigure}
\usepackage{dsfont}
\usepackage{amssymb}
\usepackage{setspace}
\usepackage{graphicx}
\usepackage{float}
\usepackage{pgfplots}
\usepackage{pgfplotstable}
\pgfplotsset{compat=1.17}
\newtheorem{remark}{Remark}
\newtheorem{lemma}{Lemma}

\newtheorem{proposition}{Proposition}

\newtheorem{proof}{\textbf{Proof}}

\def\begcen{\begin{center}}
\def\endcen{\end{center}}
\newcommand{\col}{\mbox{col}}
\newcommand{\rank}{\mbox{rank }}

\def\bfe{{\bf e}}

\def\calh{{\mathcal H}}
\def\cali{{\mathcal I}}

\def\calf{{\mathcal F}}
\def\cala{{\mathcal A}}

\def\call{{\mathcal L}}
\def\calb{{\mathcal B}}
\def\calc{{\mathcal C}}

\def\hal{\frac{1}{2}}

\def\L2e{{\mathcal L}_{2e}}

\def\rea{\mathbb{R}}

\def\l2{{\mathcal L}_2}
\def\l2e{{\cal L}_{2e}}

\def\rea{\mathbb{R}}

\def\begequarr{\begin{eqnarray}}
\def\endequarr{\end{eqnarray}}
\def\begequarrs{\begin{eqnarray*}}
\def\endequarrs{\end{eqnarray*}}
\def\begarr{\begin{array}}
\def\endarr{\end{array}}
\def\begequ{\begin{equation}}
\def\endequ{\end{equation}}
\def\begpro{\begin{proposition}}
\def\endpro{\end{proposition}}
\def\begproo{\begin{proof}}
\def\endproo{\end{proof}}
\def\beglem{\begin{lemma}}
\def\endlem{\end{lemma}}
\def\lab{\label}
\def\begdes{\begin{description}}
\def\enddes{\end{description}}
\def\begenu{\begin{enumerate}}
\def\begite{\begin{itemize}}
\def\endite{\end{itemize}}
\def\endenu{\end{enumerate}}

\def\lef[{\begin{array}}
\def\rig]{\end{array}}

\def\begcen{\begin{center}}
\def\endcen{\end{center}}
\def\begrem{\begin{remark}\rm}
\def\endrem{\end{remark}}

\def\begali#1{\begin{align}{#1}\end{align}}
\def\begalis#1{\begin{align*}{#1}\end{align*}}

\def\begmat#1{\begin{bmatrix}#1\end{bmatrix}}

\def\bfe{{\bf e}}

\def\intnum{\mathbb{Z}}

\def\begsubequ{\begin{subequations}}
	\def\endsubequ{\end{subequations}}
\usepackage{color}

\DeclareMathOperator{\detQ}{\det\{Q(t)\}}

\newcommand{\R}{\mathbb{R}}

\newcommand{\block}[4]{\begin{bmatrix}#1 & #2 \\ #3 & #4\end{bmatrix}}
\newcommand{\bigblock}[4]{\left[\begin{array}{cc}#1 & #2 \\ #3 & #4\end{array}\right]}

\def\hal{{1 \over 2}}

\graphicspath{{Figures/}}

\begin{document}
\begin{frontmatter}

\title{An Algebraic Observer for State-Affine Systems\thanksref{footnoteinfo}} 

\thanks[footnoteinfo]{
Corresponding author: Anton Pyrkin (pyrkin@itmo.ru).}

\author[ITAM]{Romeo Ortega}
\author[ITMO]{Alexey Bobtsov},
\author[HDU,ITMO]{Anton Pyrkin},
\author[ITMO]{Olga Zarina},

\address[ITMO]{Department of Control Systems and Robotics, ITMO University, Kronverkskiy av. 49, Saint Petersburg, 197178, Russia}
\address[ITAM]{Departamento Acad\'emico de Ingenier\'ia El\'ectrica y Electr\'onica, ITAM,  Progreso Tizap\'an, R\'io Hondo 1, Ciudad de M\'exico, 01080, M\'exico}
\address[HDU]{HDU-ITMO Joint Institute, Hangzhou Dianzi University, Hangzhou, 310005, China}

\begin{abstract}
We present a new {\em systematic} method to design an {\em algebraic swapping lemma observer}  \citep{BOBetal_tac26_aslo} for nonlinear systems which are affine in the state. Four are the main features of the new observer: (i) Far superior transient performance compared to standard asymptotically convergent observers; (ii) structural simplicity---its construction relies solely on basic linear filtering and a single matrix inversion; (iii) applicability to non-uniformly-completely-observable systems, for which classical methods are not applicable; (iv) straightforward extension to the cases of unknown time-varying parameters and measurement delay. The core idea exploited in the paper is to combine the construction of a generalized parameter estimation-based observer \citep{ORTetal_aut21_gpebo} with  a simple filtering technique to derive the algebraic observer. The resulting design is shown to apply to a class of state-affine systems satisfying a {\em weak generic condition}. Moreover, it is established that for time-invariant systems, this condition is {\em always} satisfied globally.
 
\end{abstract}


\end{frontmatter}
%
\section{Introduction and Problem Formulation}
\lab{sec1}
A class of nonlinear systems for which the problem of designing a state observer is well-known, are the so-called {\it state-affine} systems, whose dynamics is described by \cite[Equation 3.1]{BERbook}:
\begali{
	\nonumber
	\dot x &=\cala(y,u)x+\calb(y,u)\\
	\lab{staaff}
	y & = h(x,u),
}
where $x(t) \in \rea^n$ is the {\it unmeasurable} state, $y(t)  \in \rea^p$ denotes the measured output signal and $u(t) \in \rea^m$ is the control input. When the read-out map is linear in $x$, that is, $h(x,u)=\calc x$ the most famous observer used for this kind of systems is the Kalman-Bucy's Filter (KBF) introduced in \citep{KALBUC} for linear time-varying (LTV) systems---which are obtained evaluating  the matrices $\cala,\calb$  and $\calc$ of the system above along the input and output trajectories. That is, rewriting the system \eqref{staaff} as the LTV system
\begali{
	\nonumber
	\dot x &=A(t)x+B(t)u\\
	y &=C x,
	\lab{sys}
}
with $A(t) \in \rea^{n \times n},B(t) \in \rea^{n \times m}$ and $C \in \rea^{p \times n}$ are defined via
$$
A(t):=\cala(y(t),u(t)),\;B(t):=\calb(y(t),u(t)),\;C:=\calc.
$$

It is well-known that to ensure global convergence for this observer it is necessary to impose very strong excitation assumptions on the system, namely, {\it uniform complete observability} (UCO) of the pair $(A(t),B(t))$ \citep{BERbook, KALBUC}. In a recent paper \citep{WANORTBOB} it was shown that state observation of LTV systems is possible imposing only the (necessary) assumption of {\it observability}. For, it is proposed to use the recently reported {\it generalized parameter estimation-based observers} (GPEBO) \citep{ORTetal_aut21_gpebo}. The main feature of GPEBO is that the problem of state observation is recast as a problem of {\it parameter} estimation, namely of the systems initial conditions. This approach has proven to be very successful for the state observation of state-affine systems, and many extensions and practical applications to the method have been reported \citep{BEZetal,BOBetalaut21,BOBetalijc, PYRetal,ROMORT}.

In the recent publication \citep{BOBetal_tac26_aslo} a novel approach to the problem of state observation was reported. The main feature of these new observers is that we center our attention in the design of {\em algebraic observers}---the qualifier ``algebraic" meaning that we want to obtain an {\em algebraic} relation between the system state and filtered versions of the systems inputs and outputs, which holds true {\em for all $t \geq 0$}. The algebraic property should be contrasted with the usual procedure of designing a dynamical system whose output (the state estimate) {\em asymptotically} (or in fixed/finite time) converges to the systems state. As shown in \citep{BOBetal_tac26_aslo} a key component in the design of this algebraic observer is the application of the {\em Swapping Lemma}  \cite[Lemma 3.6.5]{SASBODbook}, hence we give to the new observer the name Algebraic Swapping Lemma Observer (ASLO).

A key feature of the ASLO is that, in contrast with standard asymptotically convergent observers, its transient behavior is {\em not dependent} on the prior knowledge of the systems initial state. Indeed, in classical observer the {\em initial value} of the state estimation error $\tilde x(t):=\hat x(t)-x(t)$ is clearly $\hat x(0)-x(0)$, where $\hat x(t)$ is the state of a dynamic system-based observer,  whose initial conditions are chosen by the designer. In ASLO, as the construction of the estimate is algebraic---and not via the construction of an auxiliary dynamical system---such a direct dependence is conspicuous by its absence.  

The application of ASLO for general nonlinear systems of the form $\dot x=f(x,u),\;y=h(x)$ is restricted to the case when the {\em derivative} of some of the states {\em is known}---or a new state can be constructed for which this property is verified \cite[Assumption 1]{BOBetal_tac26_aslo}. Clearly, the construction of such an ``auxiliary" state involves the solution of a partial differential equation that, as is well known, is hard to obtain. The main contribution of this paper is the proof that for the particular case of state-affine systems \eqref{staaff} it is {\em always} possible to design an ASLO, and an explicit, systematic procedure to do it is given in the paper. The main idea exploited in the paper is to combine the construction of a GPEBO \citep{ORTetal_aut21_gpebo} with a simple filtering technique to derive the ASLO.

The remainder of the paper is organized as follows. Section \ref{sec2} gives the main result.  In Section \ref{sec3} we consider the case of LTI systems. In Section \ref{sec4} we apply the method to solve the challenging problem of state estimation of systems which are {\em not UCO} or their states converge to non observable ones. Section \ref{sec5} presents some simulation results for LTI and LTV systems. In Section \ref{sec6} we present extensions of the result to the cases of systems with measurement delay, unknown parameters and switching parameters. The paper is wrapped-up with concluding remarks in  Section \ref{sec7}. The proof of some propositions, which are technically burdensome, are  given in Appendices \ref{appa} and \ref{appb}.\\

\noindent {\bf Notation.} ${\bf I}_n$ is the $n \times n$ identity matrix, ${\bf 0}_{n \times s}$ is an $n \times s$ matrix of zeros, $\bfe_j \in \rea^n$ is the $j$-th component of the Euclidean basis, and ${\bf 1}_{n}$ is an $n$-dimensional vector of ones. $\rea_+$ and $\intnum_+$ denote the positive real and integer numbers, respectively. For $q \in \intnum_+$ we define the set $\bar q:=\{1,2,\dots,q\}$. For a vector $a \in \rea^n$, we denote its Euclidean norm by $|a|$ and for a matrix $A \in \rea^{n \times m}$ its Euclidean matrix norm is denoted $\|A\|$. To simplify the notation, the time argument in the signals is avoided when clear from the context. The action of an LTI filter $\calf(p) \in \rea(p)$ on a signal $w(t)$ is denoted as $\calf(p)[w]$, where $p^n[w]:=\frac{d^n w(t)}{dt^n}$.
%
\section{Main Result}
\lab{sec2}
%
In this section we present the new construction of the ASLO for state-affine systems proceeding from the design of a GPEBO and adding a required dynamic extension. 

\begpro \em
\lab{pro1}
Consider the LTV system \eqref{sys} with $A(t) \in \rea^{n \times n},B(t) \in \rea^{n \times m}$ and $C \in \rea^{p \times n}$ {\em known}, with $C$ {\em constant}, and the unforced system $\dot x =A(t)x$ stable. Assume the ratio $\frac{n}{p}$ is a rational number.

Define the dynamic extension
\begali{
	\nonumber
	\dot \xi & = A(t) \xi + B(t)u,\;\xi(0)=\xi_0 \in \rea^n\\
	\lab{dotphi}
	\dot \Phi &= A(t) \Phi,\;\Phi(0)=I_n
}
and the filtered signals
\begsubequ
\lab{filsig}
\begali{
	\lab{phif}
	\dot \Phi^F_i &=-\lambda_i \Phi^F_i  + \lambda_i \Phi,\quad \Phi^F_i (0) = {\bf 0}_{n \times n}\\
	\lab{xif}
	\dot \xi^F_i &=-\lambda_i \xi^F_i + \lambda_i \xi,\quad \xi^F_i (0)={\bf 0}_{n \times 1}\\
	\lab{yf}
	\dot y^F_i&=-\lambda_i y^F_i + \lambda_i y,\quad y^F_i(0)={\bf 0}_{p \times n},
}
\endsubequ
where $i \in \overline{n-1}$, $\lambda_i>0$ and $\lambda_i \neq \lambda_j,\;\forall i \neq j$, which are {\em tuning parameters}. Define the matrices
\begali{
	\lab{yq}
	Y&:=\begmat{y - C \xi \\ y^F_1-C \xi^F_1 \\ \vdots \\ y^F_{n-1}-C \xi^F_{n-1}}\in \rea^{np},\; \nonumber \\
	Q&:=\begmat{C \Phi  \\ C \Phi^F_1 \\ \vdots \\ C \Phi^F_{n-1}} \in \rea^{np \times n}.
}
\begenu
\item[{\bf P1}]  {\em Assume} that $\det\{Q^\top(t)Q(t)\}$ is {\em non-zero} almost everywhere, with zeros only at isolated points. Define the set $\cali \subset (0,\infty]$ as the set of times when  $Q^\top(t)Q(t)$ has {\em full rank}. For $t \in \cali$ define the signal
\begequ
\lab{xaslo}
\hat x(t) :=\xi(t)-\Phi(t) Q^{\dagger}(t)Y(t).
\endequ
Then, 
$$
\hat x(t)=x(t),\quad \forall t \in \cali.
$$
That is, \eqref{xaslo} is an ASLO for the system.

\item[{\bf P2}] All signals remain bounded for all control signals $u(t)$ that ensure the state trajectories remain {\em bounded}.
\endenu
\endpro

\begin{proof}\em
	The proof of {\bf P1} proceeds first constructing a GPEBO for the system \eqref{sys}, including the associated linear regression equation (LRE). Then, create an augmented  system---piling up the original LRE and its filtered version---and compute explicitly the unknown parameter of GPEBO. The last step is to give the final, {\em algebraic} expression of the state replacing the parameter vector in the definition of the state given by GPEBO.
	
	Defining the signal
	\begequ
	\lab{e}
	e:=\xi - x,
	\endequ
	we have that $\dot e = A(t)e$, as shown in \citep{ORTetal_aut21_gpebo} this implies:
	$$
	e = \Phi \theta,
	$$
	where $\theta \in \rea^n$ is a vector of {\em unknown} parameters. Moreover, replacing the equation above in \eqref{e} we have that 
	\begequ
	\lab{x}
	x  = \xi - \Phi \theta,
	\endequ
	Multiplying \eqref{x} by $C$ we get the LRE\footnote{As explained in \citep{ORTetal_aut21_gpebo}, the design of GPEBO is completed adding an estimator for $\theta$, and replacing it in \eqref{x}.}
	\begequ
	\lab{ygpebo}
	y - C \xi = C \Phi \theta
	\endequ
	
	Now, applying the LTI filters $\frac{\lambda_i}{p+\lambda_i}$ to \eqref{ygpebo}, and taking into account the dynamic equation above, we get a set of filtered LREs
	\begequ
	\lab{yfgpebo}
	y^F_i-C \xi^F_i=C \Phi^F_i \theta.
	\endequ
	
	Piling up \eqref{ygpebo} and the $q$ versions of \eqref{yfgpebo} we get the extended LRE
	$$
	Y=Q\theta,
	$$
	
	with $Y$ and $Q$ defined in \eqref{yq}. Under the assumption of full rank $Q$ we can write
	$$
	\theta=Q^{\dagger}Y,
	$$
	that, replaced in \eqref{x} yields \eqref{xaslo} completing the proof of {\bf P1}.\\
	
	The proof of {\bf P2} follows from \eqref{dotphi}, the assumption of stability of the autonomous system and the condition imposed on the control signal $u(t)$.
\end{proof}

\begrem
\lab{rem1}
We have constructed the key matrix $Q$ applying a simple first order LTI filter $\frac{\lambda_i}{p + \lambda_i}$ to the LRE \eqref{ygpebo}. It is clear that we can achieve a similar extension applying {\em any linear operator} $\calh: \call_\infty^p \mapsto  \call_\infty^{(n-1)p}$. As shown in \citep{ORTetal_tac21_drem} the use of different operators conveys special properties to the subsequent operations carried out with the matrix $Q$. In our particular case we would be interested in ensuring it is {\em full rank}.
\endrem

\begrem
\lab{rem2}
The assumption of $\frac{n}{p}$ a rational number, is made to simplify the notation. If this is not the case, we must introduce as many filters as needed to make $Q$ a tall matrix to ensure the pseudo-inverse matrix is the {\em left} inverse.
\endrem
%
\section{Case of LTI Systems}
\lab{sec3}
%
In this section we consider the particular case of LTI systems.  Interestingly, for this case we can prove that $\detQ \neq 0$ under precise {\em generic} conditions on the filters coefficients. For the sake of ease of presentation we consider the case of single output systems, that is, $p=1$. The extension for the multi-output case follows {\em mutatis mutandi}.

\begpro \em
\lab{pro2}
Consider the case of $n$-th order, single output, LTI {\em observable} systems. Let $\sigma\{A\}$ be the set of eigenvalues of $A$, denoted $\mu_k \in \mathbb{C}$. Define the matrix $Q(t)$ given in Proposition \ref{pro1}.
\begenu
\item[{\bf S1}] Assume all eigenvalues of $A$ are {\em real}. Then, the following equivalence holds true
\begin{align*}
	\det &\{Q(t)\} \neq 0,\;\forall t > 0 \quad \Longleftrightarrow \\
	 \quad &\sigma\{A\} \cap \{-\lambda_1,-\lambda_2,\dots,-\lambda_{n-1}\} = \varnothing.
\end{align*}

\item[{\bf S2}] If some of the eigenvalues are {\em complex},\footnote{With possibly repeated eigenvalues---here and in {\bf S1}.} that is
$$
\exists k \in \bar{n}\;|\;\mu_k=\alpha_k+j\beta_k,\;\alpha_k,\beta_k \in \rea,\;\beta_k>0.
$$
Then, there exists a set of {\em isolated} time instants $\{t_j\}_{j=1}^\infty$ when 
$$
\det\{Q(t_j)\}=0.
$$
\endenu
\endpro

\begproo \em
The proof of the proposition is given in Appendix \ref{appa}.
\endproo

\begrem
\lab{rem3}
It is clear from the description above that the function $\det\{Q(t)\}$ satisfies the conditions of Proposition \ref{pro1}.
\endrem
%
\section{Two non-UCO Examples}
\lab{sec4}
%
In this section we design an ASLO for two examples whose state {\em cannot be observed} with standard techniques. The first one is the dynamics of a non-holonomic vehicle which, as is well-known \citep{WANORTBOB}, does not satisfy the UCO property, therefore the KBF cannot be used to estimate the state. The second example is also a non-UCO system. 
\subsection{State Estimation of Non-Holonomic Vehicles}
\lab{subsec41}
Consider a non-holonomic vehicle of the form \citep{BERbook}
\begin{equation}
	\label{eq:nonHV}
	\begin{array}{rcl}
		\dot z_1 &=& u_1\cos(z_3)\,\\
		\dot z_2 &=& u_1\sin(z_3)\,\\
		\dot z_3 &=& u_1u_2
	\end{array}
\end{equation}
where the inputs $u_1$ and $u_2$ represent the norm of the vehicle velocity and the orientation of the front steering wheels, respectively, and the measurable outputs are $y=\col(z_1,z_2)$. We follow the linearization approach of \cite[Subsection 6.3.3]{BERbook} by defining
\[
x_1=z_1\,, \quad x_2 = z_2\,,\quad x_3 = \cos(z_3)\,,\quad x_4 = \sin(z_3)\,.
\]
In this way, the system \eqref{eq:nonHV} is transformed into a state-affine form that can be rewritten as an LTV system,  with the definitions
\begin{align}
\lab{nonholsysA}
A(t) &= \begin{bmatrix}
	0 & 0 & u_1(t) & 0 \\
	0 & 0 & 0 & u_1(t) \\
	0 & 0 & 0 & -u_1(t)u_2(t) \\
	0 & 0 & u_1(t)u_2(t) & 0
\end{bmatrix}\,, \\ \quad
C^\top &= \begin{bmatrix}
	1 & 0 \\
	0 & 1 \\
	0 & 0 \\
	0 & 0
\end{bmatrix}, \\ 
\quad B(t)&=0. \lab{nonholsysB}
\end{align}
Let
\begequ
\lab{u1u2}
u_1=e^{-t}\,,\quad u_2 = \left\{\begin{array}{l}2-2t\,,\qquad t\leq 1 \\ 0 \,,\qquad \qquad t> 1 \end{array}\right.\,.
\endequ
It is easy to show that the resulting LTV system is observable but {\em not UCO}---therefore, it is not amenable for the application of the Kalman Filter for the estimation of the state. 

The lemma below proves that it is possible to apply the ASLO observer of Proposition \ref{pro1}

\beglem
\lab{lem1}\em
Consider the LTV system \eqref{sys} with \eqref{nonholsysA}-\eqref{nonholsysB} and \eqref{u1u2}. For all values of the filter constant $\lambda>0$, the matrix $Q(t)$ constructed in Proposition \ref{pro1} satisfies that $\det\{Q(t)\}$ is {\em non-zero} almost everywhere, with zeros only at isolated points.
\endlem
\begproo \em
The proof is given in Appendix \ref{appb}.
\endproo

\begin{figure}[ht]
	\centering
	\includegraphics[width=1\linewidth]{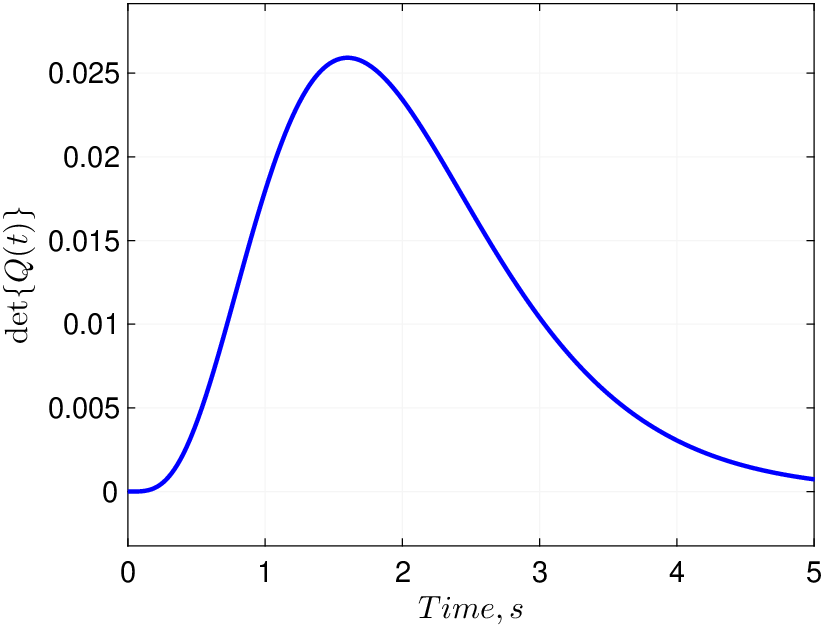}
	\caption{Signal $\detQ$}
	\label{fig1}
\end{figure}

\begin{figure}[ht]
	\centering
	\includegraphics[width=1\linewidth]{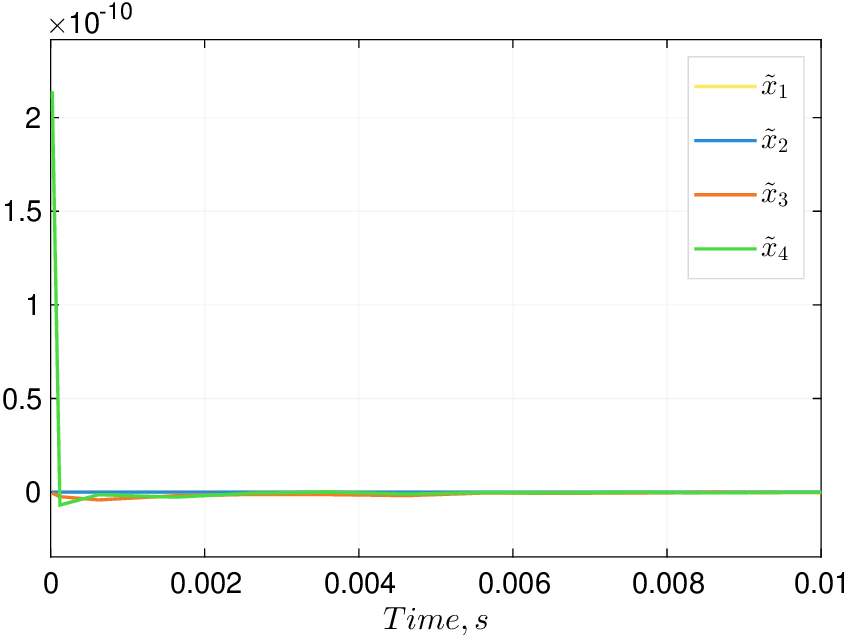}
	\caption{State estimation errors}
	\label{fig2}
\end{figure}

We apply the ASLO of Proposition \ref{pro1} to estimate the state $x(t)$, and the resulting simulations using $\lambda=1$ are given in Figs. \ref{fig1} and \ref{fig2}.  It can be seen from Fig.  \ref{fig1} that $\detQ$ satisfies the conditions of the lemma---being zero only at $t=0$ and $t=\infty$. From Fig.  \ref{fig2} we see that the observed states $\hat x(t)$ follow the actual states $x(t)$ {\em almost immediately}---please, notice the time scale.

\begin{figure}[ht]
	\centering
	\includegraphics[width=1\linewidth]{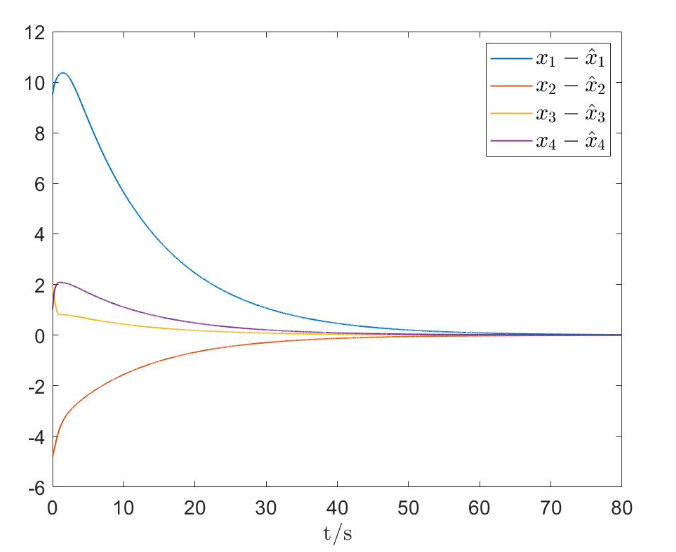}
	\caption{State estimation errors $\hat x(t)-x(t)$ of the GPEBO reported in  \citep{WANORTBOB}}
	\label{figlei}
\end{figure}

The state observation of the system given above has been addressed in \citep{WANORTBOB} using the GPEBO proposed in that paper. It is very revealing to compare the transient of the state error signals for the ASLO shown in Fig. \ref{fig2} with the ones reported in  Fig. 4 in that paper, which are repeated here in Fig. \ref{figlei} for ease of reference.  Contrary to the almost {\em instantaneous tracking} of ASLO the GPEBO takes more than $40s$ to converge. This, unquestionably reveals the superiority of the new ASLO. 

\begin{figure}[ht]
	\centering
	\includegraphics[width=1\linewidth]{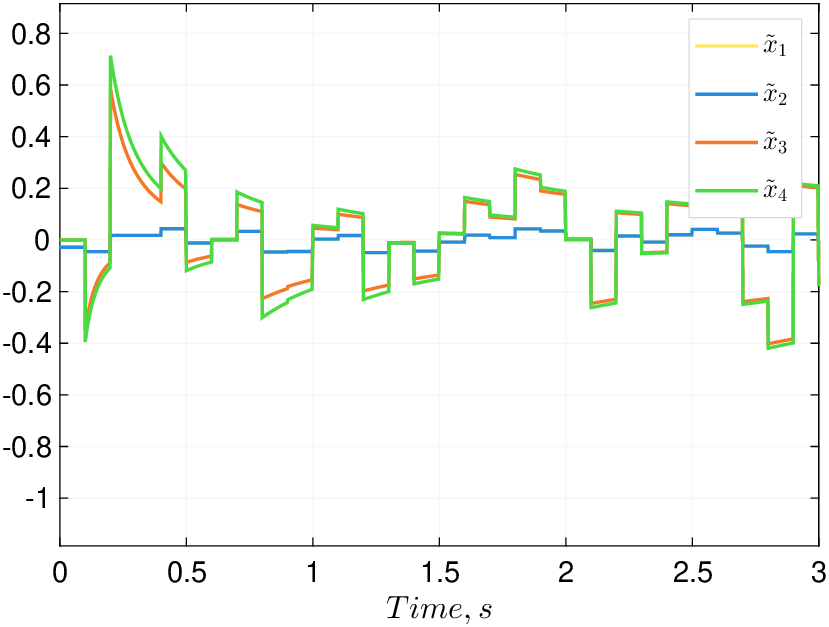}
	\caption{State estimation errors $\hat x(t)-x(t)$ under measurement noises}
	\label{fig3}
\end{figure}

It should be mentioned that to study the robustness of the GPEBO of   \citep{WANORTBOB}, the case with measurement noises which are generated by the block ``Uniform Random Number" of Simulink was also simulated and only small  state estimation errors are observed, showing robustness to measurement noise---see \cite[Fig. 5]{WANORTBOB}. On the other hand, repeating this noisy simulation for the ASLO resulted in a serious degradation of the performance as shown in Fig. \ref{fig3}. Current research is under way to try to improve this drawback of our ASLO design. 

As a comparison, in \citep{WANORTBOB} the KBF of \cite[Theorem 3.3]{BERbook} was also used derive the state estimate. The  simulation result is given in  \cite[Fig. 6]{WANORTBOB}, where it is seen that the state estimation errors {\em do not converge} to zero. This fact is not surprising since there is no guarantee of asymptotic state estimation using KBF for systems that are not UCO.

\subsection{A simple non-UCO system}
\lab{subsec42}
Consider the two-dimensional LTV system \eqref{sys},  with the definitions
\begequ
\lab{nonobssys}
A(t) = \begin{bmatrix}
	0 & e^{-kt}\\
	0 & 0 
\end{bmatrix}\,,\quad
C = \begin{bmatrix}
	1 & 0
\end{bmatrix},\quad B(t)=0.
\endequ
with $k>0$. The state-transition matrix $\Phi(t,\tau)$, is given by
\[
\Phi(t,\tau) =
\begin{bmatrix}
	1 & \dfrac{1}{k}(e^{-k\tau} - e^{-kt}) \\[8pt]
	0 & 1
\end{bmatrix}.
\]
Consequently, the principal matrix
\[
\Phi(t):=\Phi(t,0) =
\begin{bmatrix}
	1 & \dfrac{1}{k}(1 - e^{-kt}) \\[6pt]
	0 & 1
\end{bmatrix}.
\]
Some simple calculations show that the observability Gramian on $[0,T]$ is
\[
W_o(0,T) =
\begin{bmatrix}
	T & \dfrac{kT - \eta}{k^2} \\[10pt]
	\dfrac{kT - \eta}{k^2} &
	\dfrac{2k^2 T - 4k\eta + (1 - e^{-2kT})}{2k^3}
\end{bmatrix},
\]
with $\eta=1-e^{-kT}$, whose determinant is
\[
\det \{W_o(0,T)\} =
\frac{1}{2k^4}\eta^2 - 2kT\,e^{-kT}.
\]
At $T\to 0^+$, $\det W_o(0,T)\to 0$; hence the system is \emph{not UCO} and the KBF cannot be applied for the estimation of the state. On the other hand, the point-wise observability matrix yields
\[
\mathcal{O}(t) =
\begin{bmatrix}
	C \\ C(t)A(t) + \dot{C}(t)
\end{bmatrix}
=
\begin{bmatrix}
	1 & 0 \\ 0 & e^{-kt}
\end{bmatrix}
\]
which has full rank for all $t>0$, hence the system is observable and the ASLO is applicable.

The filtered matrix equation, with $\lambda \neq k$ and initial condition $\Phi^F(0) ={\bf 0}_{2 \times 2}$, yields
\[
\Phi^F(t) =
\begin{bmatrix}
	1 - e^{-\lambda t} & \displaystyle \frac{1 - e^{-\lambda t}}{k}
	- \frac{\lambda\left(e^{-kt} - e^{-\lambda t}\right)}{k(\lambda - k)} \\[10pt]
	0 & 1 - e^{-\lambda t}
\end{bmatrix}.
\]
Consequently
\begin{align}
\nonumber
Q(t) &=
\begin{bmatrix}
	C\,\Phi(t) \\[3pt]
	C\,\Phi^F(t)
\end{bmatrix} = \\
&=\begin{bmatrix}
	1 & \dfrac{1 - e^{-kt}}{k} \\[8pt]
	1 - e^{-\lambda t} & \displaystyle \frac{1 - e^{-\lambda t}}{k}
	- \frac{\lambda\left(e^{-kt} - e^{-\lambda t}\right)}{k(\lambda - k)}
\end{bmatrix}.
\end{align}
The determinant is
\[
\det \{Q(t)\} =  \frac{e^{-kt}(1 - e^{-\lambda t})}{k}
- \frac{\lambda\left(e^{-kt} - e^{-\lambda t}\right)}{k(\lambda - k)}.
\]
Therefore, we conclude that 
$$
\rank\{Q(t)\} = 2,\; \infty > t > 0,
$$
verifying the conditions of Proposition \ref{pro1}.

\begin{figure}[ht]
	\centering
	\includegraphics[width=1\linewidth]{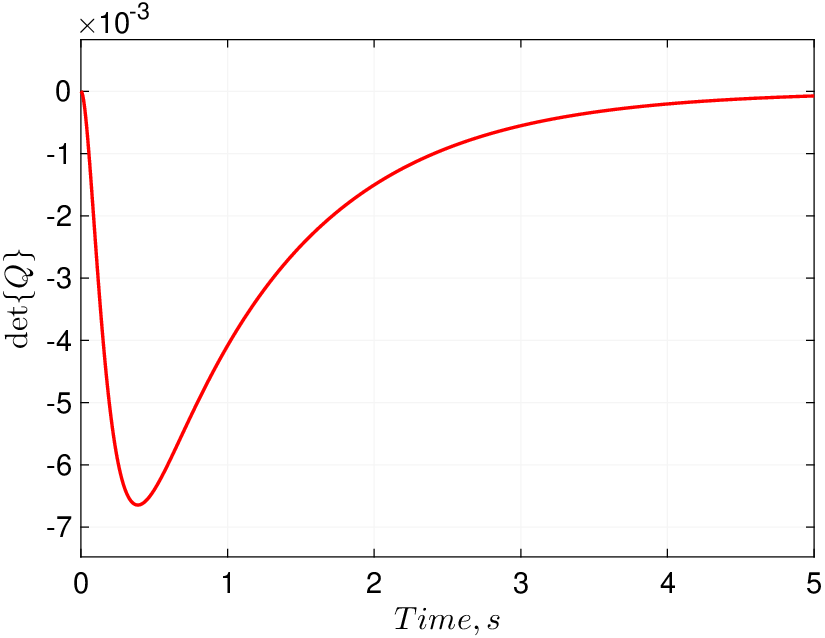}
	\caption{Signal $\detQ$}
	\label{fig5}
\end{figure}

\begin{figure}[ht]
	\centering
	\includegraphics[width=1\linewidth]{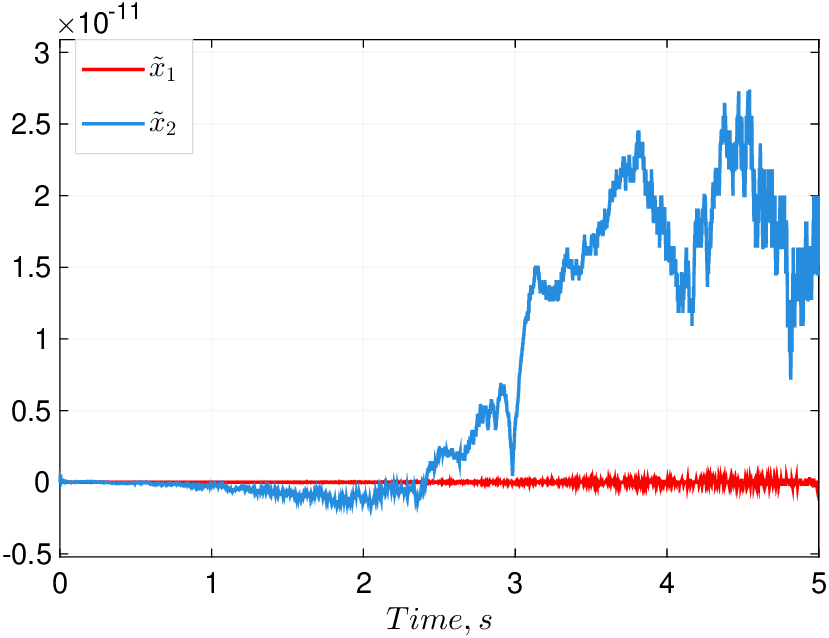}
	\caption{State estimation errors}
	\label{fig6}
\end{figure}

In \ref{fig5} we plot $\det\{Q(t)\}$ and in Fig \ref{fig6} the state estimation errors. We believe the wiggly behavior of the latter signals is due to numerical instability---please, notice the very small amplitude scale.

\section{Simulation Results for Linear Systems}
\lab{sec5}
%
Simulations were performed  for a third order LTI system and a second order UCO LTV system. For these systems, besides an ASLO, a Luenberger observer and a KBF were designed, respectively. The objective of this simulation is to compare the performance a a classical observer with our ASLO. 

\subsection{Third order LTI system}
\lab{subsec51}
In this subsection we present simulation results for a third order stable LTI system. The parameters of the system were chosen as follows:
\begin{equation}
	A = \begin{bmatrix}
		-4 &1 &2\\
		-1 &-3 &1 \\
		2 &-1 &-5
	\end{bmatrix}, B = \begin{bmatrix}
		1 \\
		-3 \\
		2
	\end{bmatrix}, C = \begin{bmatrix}
		1 \\ 0 \\ 0
	\end{bmatrix}.
\end{equation}
The initial condition of the state are $x(0)=\col(1,2,2)$.
The control signal is $u=5\sin(10t)$. It can be easily shown that the system is observable. 

\begin{figure}[H]
	\centering
	\includegraphics [width=1\linewidth]{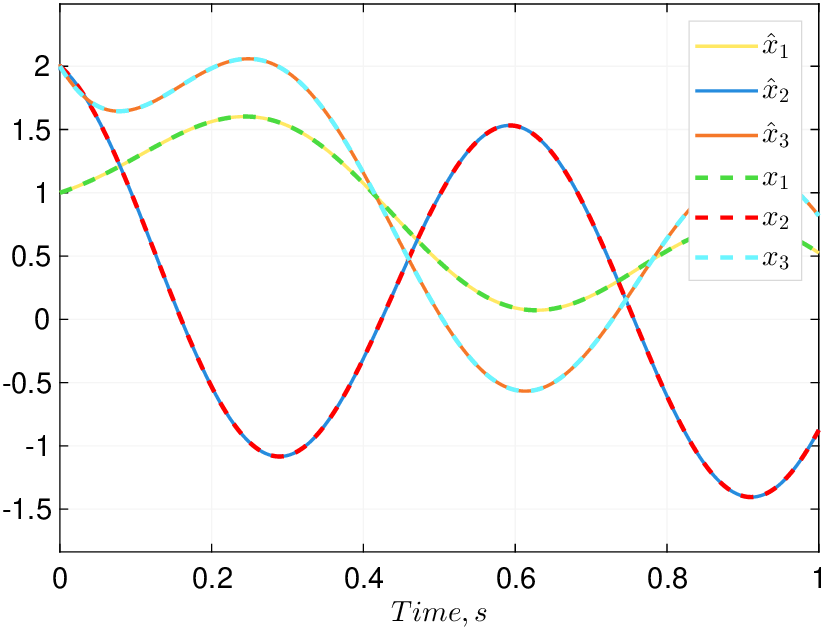}
	\caption{Transient of the state vector $x(t)$ and state vector estimate $\hat x(t)$. }
	\label{fig7}
\end{figure}

\begin{figure}[H]
	\centering
	\includegraphics [width=1\linewidth]{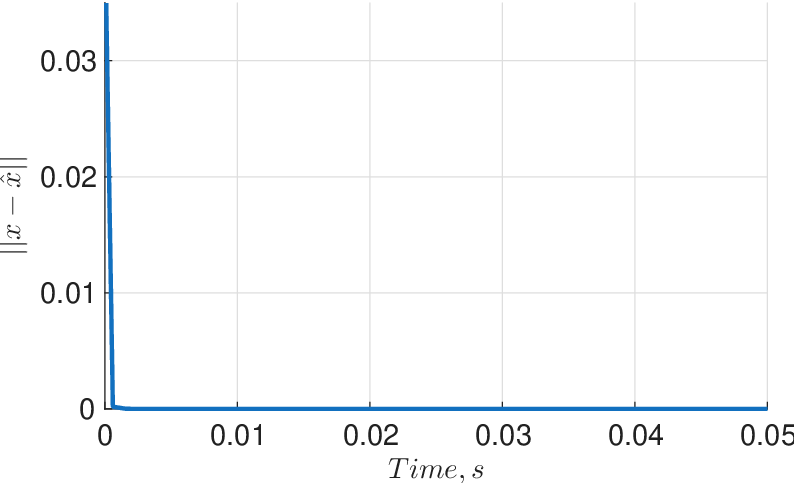}
	\caption{Transient of the estimation error norm $|x(t)-\hat x(t)|$. }
	\label{fig7.1}
\end{figure}

\begin{figure}[H]
	\centering
	\includegraphics [width=1\linewidth]{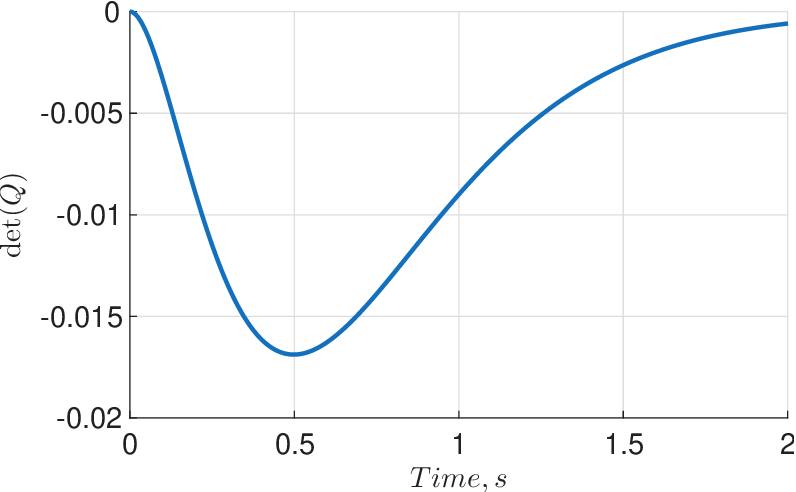}
	\caption{Transient of $\detQ$}
	\label{fig8}
\end{figure}

The results for the ASLO with $\lambda_1=0.1$ and $\lambda_2=0.01$ are shown in Fig. \ref{fig7}, \ref{fig7.1} and \ref{fig8}. As expected the behavior is extraordinary. 

The poles of the system are 
\[
p_1 = -5.88, \quad p_2 = -3.65, \quad p_3 = -2.47.
\]
A Luenberger observer of the classical form
\[
\dot{\hat{x}} = A\hat{x} + Bu + L(y - C\hat{x})
\]
was designed for this system for various choices of the observer poles.
\begenu
\item Fast poles $\{-10, -11, -12\}$, with $L= \begin{bmatrix} 21 \\ -819 \\ 484 \end{bmatrix}$.
\item Slower poles $\{-5, -6, -7\}$, with $L=\begin{bmatrix} 6 \\-24 \\ 19 \end{bmatrix}$.
\item Faster poles $\{-20, -21, -22\}$, with $L=\begin{bmatrix} 51 \\-10659\\5764 \end{bmatrix}$.
\item Complex poles $\{-5\pm 2j, -6\}$, with $L=\begin{bmatrix} 4\\-30\\21  \end{bmatrix}$.
\endenu

The transient behavior of the fours designs is shown in Fig. \ref{fig9}. It is clear that performance of the ASLO is significantly superior to that of the Luenberger design. It is clearly observed that to achieve a reduced settling time an inadmissibly large pick---of 150 times the initial error---in the transient errors is observed! A reduced order Luenberger observer was also simulated achieving basically the same performance.

\begin{figure}[h!]
	\centering
	\includegraphics[width=0.44\textwidth]{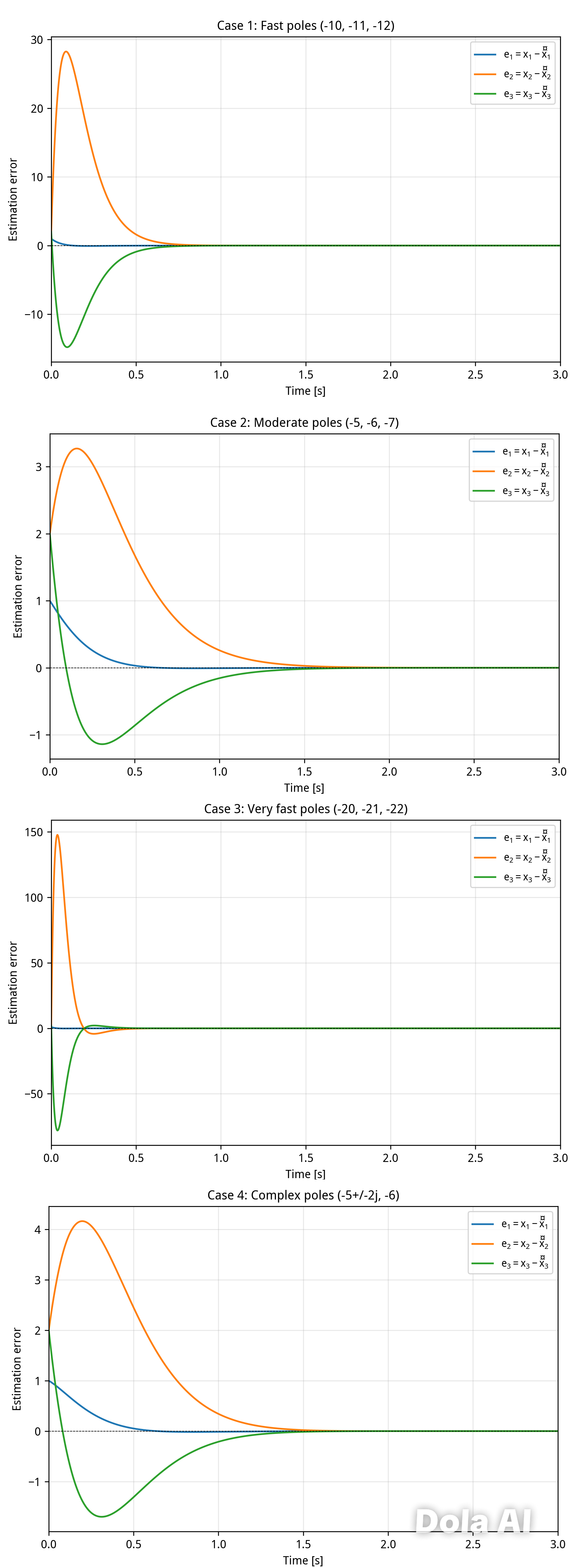}
	\caption{State estimation errors of the Luenberger observer for the four desired pole locations.}
	\lab{fig9}
\end{figure}

\subsection{Second order UCO LTV system}
\lab{subsec52}
Consider the system matrices
\[
A(t) = \begin{bmatrix}0 & 2+\sin t \\ 0 & 0\end{bmatrix},\quad
C = \begin{bmatrix}1 & 0\end{bmatrix}.
\]
with $x(0)=\col(3,-1)$. To investigate if ASLO is applicable we check its observability. The state transition and the principal matrices are given as
\begin{align}
\Phi(t,\tau) &= \begin{bmatrix}
	1 & 2(t-\tau) + \cos\tau - \cos t \\
	0 & 1
\end{bmatrix}, \nonumber\\
\Phi(t,0) &= \begin{bmatrix}
	1 & 2t + 1 - \cos t \\
	0 & 1
\end{bmatrix} \nonumber
\end{align}
On the other hand, the observability matrix is of the form
\begin{align}
O(t,\tau) = &\begin{bmatrix}
	C\,\Phi(t,\tau) \\[4pt]
	\dfrac{\partial}{\partial t}\bigl(C\,\Phi(t,\tau)\bigr)
\end{bmatrix} 
= \nonumber \\
&\begin{bmatrix}
	1 & 2(t-\tau) + \cos\tau - \cos t \\[4pt]
	0 & 2 + \sin t
\end{bmatrix} \nonumber
\end{align}
and its determinant is
\[
\det O(t,\tau) = 2+\sin t.
\]
Hence the system is observable. We proceed, then, to design an ASLO with $\lambda=1$. For, we compute the solution of
$$
\dot{\Phi}^F(t) = -\Phi^F(t) + \Phi(t),\;\Phi^F(0) = 0.
$$
This is given by
\[
\Phi^F(t) = \begin{bmatrix} 1 - e^{-t} & 2t - 1 + \frac{3}{2} e^{-t} - \hal(\sin t + \cos t) \\ 0 & 1 - e^{-t} \end{bmatrix}
\]
The matrix $Q(t)$ takes the form
\[
Q(t) = \begin{bmatrix} 1 & 2t + 1 - \cos t \\ 1 - e^{-t} & 2t - 1 + \frac{3}{2} e^{-t} - \frac{\sin t + \cos t}{2} \end{bmatrix},
\]
with its determinant 
\[
\det Q(t) = -2 + \frac{\cos t - \sin t}{2} + e^{-t} \left(2t + \frac{5}{2} - \cos t\right).
\]

\begin{figure}[H]
	\centering
	\includegraphics [width=0.9\linewidth]{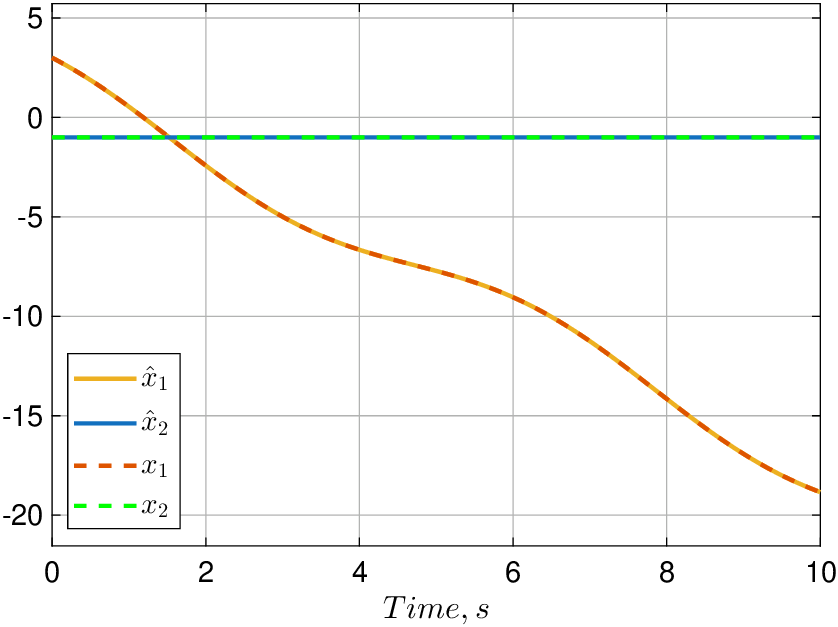}
	\caption{Transient of the state vector $x(t)$ and state vector estimate $\hat x(t)$. }
	\label{fig10}
\end{figure}

\begin{figure}[H]
	\centering
	\includegraphics [width=1\linewidth]{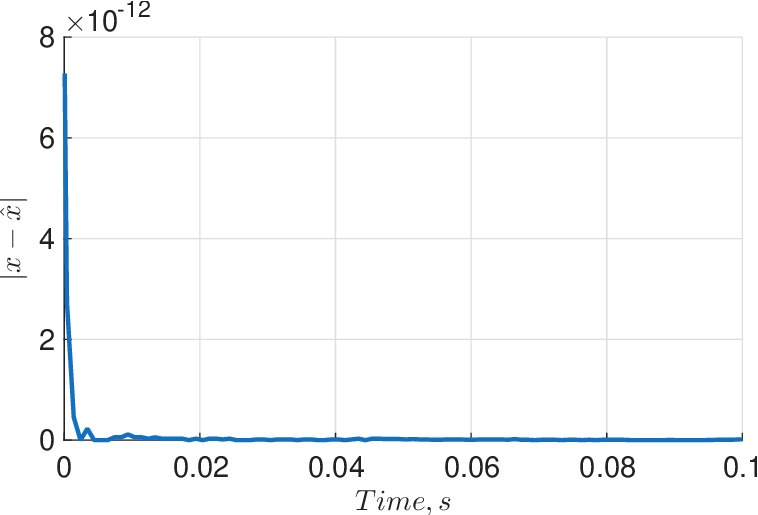}
	\caption{Transient of the estimation error norm $|x(t)-\hat x(t)|$. }
	\label{fig10.1}
\end{figure}

\begin{figure}[H]
	\centering
	\includegraphics [width=1\linewidth]{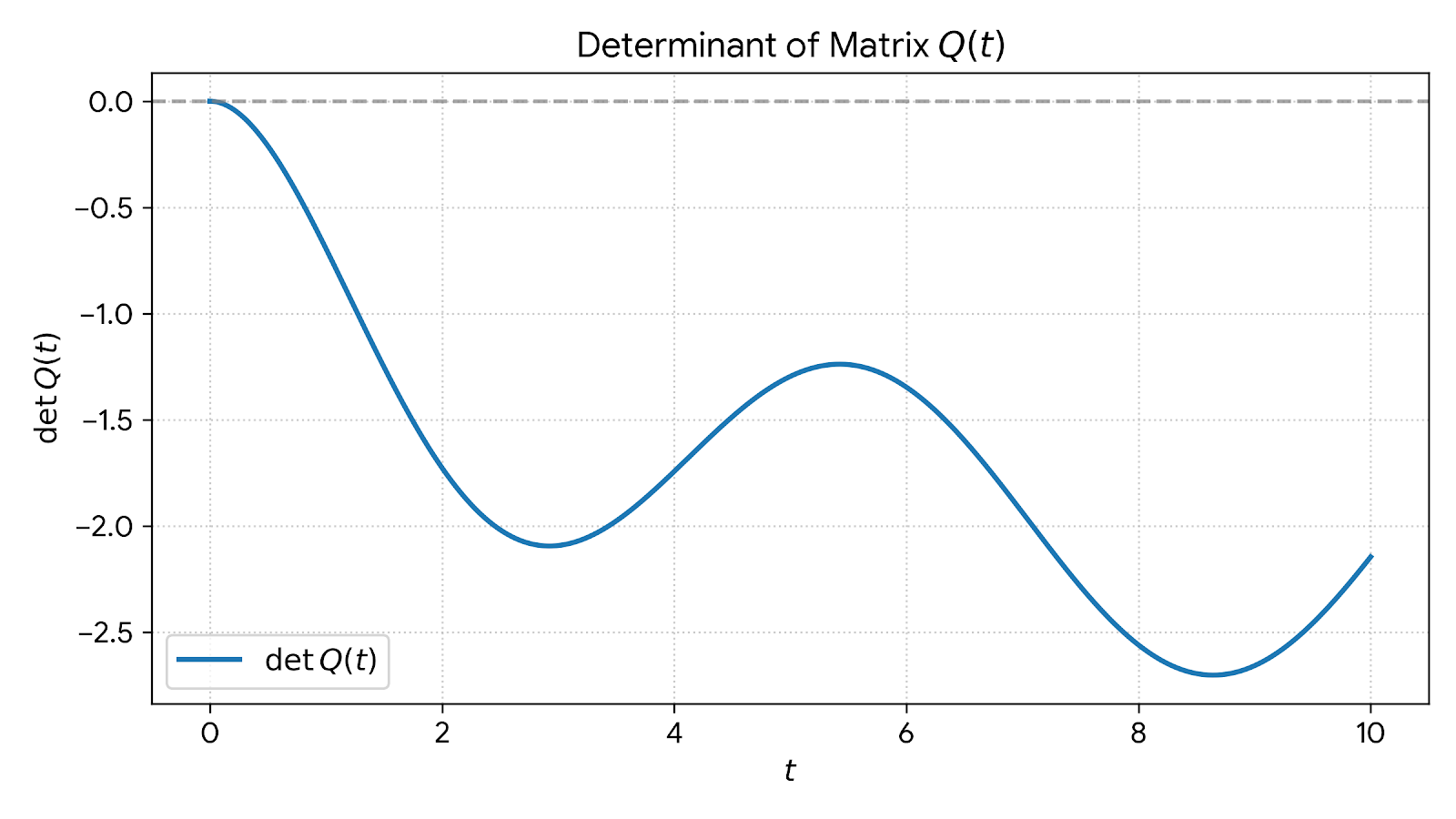}
	\caption{Transient of $\detQ$}
	\label{fig11}
\end{figure}

The transient behavior of the state estimation error and $\det\{Q(t)\}$ are shown in Figs. \ref{fig10}, \ref{fig10.1} and \ref{fig11}, respectively.

To check if we can apply the KBF to estimate the state we evaluate its UCO. Since $\det O(t,\tau) \ge 1 > 0$ uniformly in time, the system it is
UCO. Indeed, the observability Gramian
satisfies 
$$
W_o(t,t+\delta) \succeq \alpha(\delta)\,{\bf I}_2 > 0, \forall t.
$$

We proceed to design a KBF as 
\begalis{
	\dot{\hat{{x}}}(t) &= A(t)\,\hat{{x}}(t) + L(t)\bigl[y(t)-C\,\hat{{x}}(t)\bigr] \\
	\dot{P}(t) & = A(t)P(t)+P(t)A^\top(t) - {1 \over r}P(t)C^\top C P(t) \\
	L(t) & = \dfrac{1}{r}\begmat{p_{11}(t) \\ p_{12}(t)},
}
with $r>0$. A plot of the observation errors for $x(0)=(3,-1)$ and $\hat x(0)=(0,0)$ is given in Fig. 12.  We see that the settling time is about $5sec$. Again, the performance of the ASLO is {\em markedly superior} to the one of the KBF.

\begin{center}
	\begin{tikzpicture}[scale=0.7]
		\begin{axis}[
			width=12cm, height=7cm,
			xmin=0, xmax=10,
			ymin=-3.5, ymax=3.5,
			grid=major, grid style={opacity=0.3},
			xlabel={Time $t$ [s]},
			ylabel={Estimation error $\tilde{x}_i(t)$},
			legend pos=north east,
			legend style={font=\small},
			line width=0.8pt,
			]
			\addplot[blue, smooth, thick] coordinates {
				(0.0,  3.0000) (1.0,  1.5230) (2.0,  0.7940) (3.0,  0.3820) (4.0,  0.1650)
				(5.0,  0.0680) (6.0,  0.0270) (7.0,  0.0100) (8.0,  0.0035) (9.0,  0.0012)
				(10.0, 0.0004) 
			};
			\addlegendentry{$\tilde{x}_1(t) = x_1-\hat{x}_1$}
			\addplot[red, smooth, thick] coordinates {
				(0.0, -1.0000) (1.0, -0.6840) (2.0, -0.4260) (3.0, -0.2480) (4.0, -0.1350)
				(5.0, -0.0690) (6.0, -0.0340) (7.0, -0.0160) (8.0, -0.0072) (9.0, -0.0031)
				(10.0,-0.0013)
			};
			\addlegendentry{$\tilde{x}_2(t) = x_2-\hat{x}_2$}
			
			\addplot[dashed, black, opacity=0.6] coordinates {(0,0)(10,0)};
			
		\end{axis}
		\node[below, font=\small] at (current bounding box.south)
		{\textbf{Figure 12} Observation errors $\tilde{x}_1(t)$ and $\tilde{x}_2(t)$.};
	\end{tikzpicture}
\end{center}
%
\section{Some Extensions}
\lab{sec6}
%
In this section we present extensions of Proposition \ref{pro1} to the following cases.
\begenu
\item [{\bf E1}] Systems with {\em unknown time-varying parameters}. \\
\item [{\bf E2}] Measurement of the output signal with a {\em time-varying delay}.\\
\item [{\bf E3}] Systems with unknown {\em switching parameters}.
\endenu
In all cases we simply replace the derivation of the GPEBO equations reported in Proposition \ref{pro1} for the model \eqref{sys} to the one reported in a recent publication for the new model. Exact reference to the publication as well as the equations that need to be replaced is given in each subsection. To simplify the cross-referencing, we keep in the sequel the notation used in the corresponding paper and give the precise numbers of the equations.
\subsection{Systems with unknown time-varying parameters}
\lab{subsec61}
Following the derivations of \citep{BOBetalijc} we consider LTV systems described by the equations
\begalis{
\dot x &= [A(t)+\theta(t) C^\top]x+B(t)u(t)\\
y & = C^\top x
}
where the {\em unknown, time-varying} parameter $\theta(t) \in \rea^n$ satisfies the differential equation
$$
\dot \theta(t)=M(t)\theta(t),
$$
$M(t) \in \rea^{n \times n}$ {\em known} and such that $\theta(t)$ is bounded.

It is shown in \citep[Equation (11)]{BOBetalijc}  that it is possible to generate a LRE
$$
q(t)=m(t)\eta
$$
with $\eta \in \rea^{2n}$ an {\em unknown parameter} and express the state as  \citep[Equation (9)]{BOBetalijc}
$$
x(t)=\xi_1(t)+\Phi(t) \eta,
$$
where $q(t),m(t),\xi_1(t)$ and $\Phi(t)$ are {\em measurable}. 

It is clear that replacing \eqref{x} and \eqref{ygpebo} by the two equations above, we can construct the ASLO simply adding the filtered quantities.
\subsection{Delayed output measurement}
\lab{subsec62}
In \citep{BOBetalaut21} we consider the case where the measurement of the output in \eqref{sys} is of the form \citep[Equations (1),(16)]{BOBetalaut21} 
$$
y(t)=C x(t-d(t))
$$
where the {\em time-varying delay} satisfies $0 \leq d(t)\leq d_M$.

Similarly to the previous case we generate a LRE \citep[Equation (12)]{BOBetalaut21}
$$
z(t)=\Psi(t)\theta
$$
with $\theta \in \rea^{n}$ an {\em unknown parameter} and express the state as  \citep[Equation (9)]{BOBetalaut21}
$$
x(t)=\xi(t)+\Phi_a(t) \theta,
$$
where $z(t),\Psi(t),\xi_1(t)$ and $\Phi(t)$ are {\em measurable}. 
\subsection{Systems with unknown switching parameters}
\lab{subsec63}
In \citep{ROMORT} we consider switching systems of the form
\begalis{
\dot x &= A(t)x+B(t)u(t)+M(t)\theta_{\rho[k]} \\
y & = C^\top x+D(t)
}
where $A(t),B(t),M(t),C$ and $D(t)$ are {\em known} and $\theta_{\rho[k]}$ denotes the switched {\em unknown} parameter which takes values in a finite set 
$$
\theta_{\theta_{\rho[k]}} \in \Theta:\{\theta_1,\theta_2, \dots,\theta_q \},\;q\in \intnum_+.
$$
The {\em switching signal} $\theta_{\rho[k]}:\intnum_+ \mapsto \bar q$ is a {\em known} pulse-wise signal.

It is shown in \citep[Equation (8)]{ROMORT}  that it is possible to generate a LRE
$$
Y(t)=\phi(t)\eta_k
$$
with $\eta_k \in \rea^{n+m}$ an {\em unknown parameter} and express the state as  \citep[Equation (6)]{ROMORT}
$$
x(t)=\xi(t)+\Psi(t) \eta_k,
$$
where $Y(t),\phi(t),\xi(t)$ and $\Psi(t)$ are {\em measurable}. 

\begrem
\lab{rem4}
In  \citep{BEZetal} the case of systems with measurement delay and unknown time-varying parameters is considered. This corresponds to combining the cases of Subsection \ref{subsec61} and \ref{subsec62}.
\endrem

\begrem
\lab{rem5}
In \citep{PYRetal} we consider, besides unknown time-varying parameters, the presence of an external {\em additive disturbance} $\delta(t) \in \rea$ to the state equations. It is assumed that these disturbances satisfy an {\em internal model}, that is
\begalis{
\dot w(t)&=S(\rho)w(t)\\
\delta(t)&=h_\delta^\top w(t),
}
with $S$ known, but $\rho \in \rea^{n_\rho}$ and $h_\delta \in \rea^{n_w}$ {\em unknown}. A globally convergent GPEBO is designed for this observer. However, the LRE does not comply with the conditions required for the application of our technique to obtain an ASLO. Current research is underway to clear this problem.
\endrem

\begrem
\lab{rem6}
As indicated in Remark \ref{rem1} we have created our extended regressor using the simplest LTI filter, while these can be done with any {\em linear} operator. Current research is underway to investigate the potential advantages of using other operators, particularly with the objective of creating a full rank matrix $Q(t)$.
\endrem
%
\section{Concluding Remarks}
\lab{sec7}
%
The newly introduced concept of an \textit{algebraic} observer represents a major breakthrough in the field of observer design. A class of algebraic observer has been derived in an extensive series of publications, including \citep{FLIJOISIR,SIRbook}. In these works its unquestionable superiority \textit{vis-\`a-vis}  classical asymptotic observers is discussed at length. Moreover, it is argued that  the availability of algebraic observers trivializes the control design task.  Unfortunately, in all these works the authors rely on the  operation of successive {\em signal differentiation}---an operation that may be problematic for practical applications where the presence of {\em noise} is unavoidable. This undesirable feature is conspicuously  absent from our ASLO approach.

We present four key features of algebraic observers to demonstrate their superiority:

\begin{enumerate}
\item[\textbf{F1}] Far superior transient performance, stemming from the fact that 
	the estimated state signal becomes aligned with the true state almost immediately and, under 
	a generic rank condition on an algorithm-generated matrix, remains exact for all 
	subsequent time. This stands in sharp contrast to classical \textit{asymptotic 
(or finite/fixed-time)} convergent designs.
	
\item[\textbf{F2}] Structural simplicity: the construction relies solely on basic linear filtering and a single matrix inversion. These computational requirements compare favorably with the $\approx 4n^3 + 4n^2m + 2nm^2$ flops per time step required by the KBF \citep{SANetal}.
	
\item[\textbf{F3}] Applicability to {\em non-UCO systems}, or systems with trajectories converging to unobservable states, for which classical methods such as the KBF are not applicable---see Section \ref{sec4} and \citep{BOBetal_jpc26}.

\item[\textbf{F4}] As shown in Section \ref{sec6} it is straightforward to extend our design to several challenging situations, {\em e.g.}, unknown time-varying or switching parameters as well as output measurement with time-varying delay. None of these cases can be handled with existing observers.  
\end{enumerate}

As discussed in \citep{BOBetal_tac26_aslo}, the application of ASLO to general nonlinear 
systems of the form $\dot x = f(x,u)$, $y = h(x)$,
is restricted to cases where the \textit{derivative} of certain state components are
\textit{known}. Furthermore, the design proceeds on a case-by-case basis, with no 
unified systematic methodology. In contrast to this situation,  for the particular class of 
\textit{state-affine systems}, we show in this paper that an ASLO can \textit{always} be constructed and will be applicable under weak generic conditions. Moreover, we provide a simple, step-by-step, systematic design procedure that requires only the operation of linear filtering and a matrix inversion.  
%

\begcen {\bf Acknowledgments} \endcen
This paper is supported by the Ministry of Science and Higher Education of the Russian Federation (project No. FSER-2025-0002).\\

\bibliographystyle{plainnat}
\bibliography{refs} 

@book{BERbook,
	author = {P. Bernard},
	title = {Observer Design for Nonlinear Systems},
	year = {2019},
	publisher = {Springer},
	address = {Switzerland}
}

@article{BOBetal_tac26_aslo,
	author = {A. Bobtsov and J.G. Romero and R. Ortega and A. Pyrkin},
	title = {An algebraic state observer  for a class of physical systems},
	year = {2026},
	journal = {arXiv:2604.23142},
	publisher = {IEEE Transactions on Automatic Control},
	note = {(submitted)}
}

@article{BEZetal,
	author = {V. Bezzubov and A. Bobtsov and D. Efimov and R. Ortega and N. Nikolaev},
	title = {Adaptive state observation of linear time-varying systems with delayed measurements and unknown parameters},
	year = {2022},
	journal = {International Journal of Robust and Nonlinear control},
	volume = {33},
	pages = {1203--1213},
}

@article{BOBetalaut21,
	author = {A. Bobtsov and N. Nikolaev and R. Ortega and D. Efimov},
	title = {State observation of LTV systems with delayed measurements: A parameter estimation-based approach with fixed convergence time},
	year = {2021},
	journal = {Automatica},
	volume = {131},
	pages = {109674},
}

@article{BOBetalijc,
	author = {A. Bobtsov and R. Ortega and B. Yi and N. Nikolaev},
	title = {Adaptive state estimation of state-affine systems with unknown time-varying parameters},
	year = {2021},
	journal = {International Journal of Control},
	volume = {95},
	number={9},
	pages = {2460--2472}
}

@article{BOBetal_jpc26,
	author = {A. Bobtsov and A. Pyrkin and R. Ortega and J.G. Romero and D. Dochain},
	title = {An algebraic state observer  for a batch reactor},
	year = {2026},
	journal = {Journal of Process Control},
	note = {(submitted)}
}

@article{FLIJOISIR,
	author = {M. Fliess and C. Join and H. Sira-Ram?rez},
	title = {Non-linear estimation is easy},
	year = {2008},
	journal = {International Journal of Modelling, Identification and Control},
	volume = {4},
	number = {1},
	pages = {12--27}
}

@article{KALBUC,
	author = {R. Kalman and R. Bucy},
	title = {New results in linear filtering and prediction theory},
	year = {1961},
	journal = {Journal of Basic Engineering},
	volume = {108},
	pages = {83-95}
}

@article{ORTetal_aut21_gpebo,
	author = {R. Ortega and A. Bobtsov and N. Nikolaev and J. Schiffer and D. Dochain},
	title = {Generalized parameter estimation-based observers: application to power systems and chemical-biological reactors},
	year = {2021},
	journal = {Automatica},
	volume = {129},
	pages = {109635},
}

@article{ORTetal_tac21_drem,
	author = {R. Ortega and S. Aranovskiy and A. Pyrkin and A. Astolfi and A. Bobtsov},
	title = {New results on parameter estimation via dynamic regressor extension and mixing: Continuous and discrete-time cases},
	year = {2021},
	journal = {IEEE Transactions on Automatic Control},
	volume = {66},
	number = {5},
	pages = {2265--2272}
}

@article{PYRetal,
	author = {A. Pyrkin and A. Bobtsov and R. Ortega and A. Isidori},
	title = {An adaptive observer for uncertain linear time-varying systems with unknown additive perturbations},
	year = {2023},
	journal = {Automatica},
	volume = {147},
	pages = {110677},
}

@article{ROMORT,
	author = {J. G. Romero and R. Ortega},
	title = {Adaptive state observation of linear time-varying systems with switching unknown parameters: Application to gain scheduling and event-triggered control},
	year = {2023},
	journal = {International Journal on Adaptive Control and Signal Processing},
	volume = {37},
	pages = {2915--2933},
}

@INPROCEEDINGS{SANetal,
	author={H. Sandberg and J-C. Delvenne  and N. J. Newton and S. K. Mitter},
	booktitle={2014 52nd Annual Allerton Conference on Communication, Control, and Computing, Allerton}, 
	title={Thermodynamic costs in implementing Kalman-Bucy filters}, 
	year={2014},
	pages={550-555}
	}

@book{SASBODbook,
	author = {S. Sastry and M. Bodson},
	title = {Adaptive Control: Stability, Convergence and Robustness},
	year = {1989},
	publisher = {Prentice-Hall},
	address ={New Jersey}
}

@book{SIRbook,
	author = {H. Sira-Ramírez and C. García-Rodríguez and J. Cortés-Romero and A. Luviano-Juárez},
	title = {Algebraic Identification and Estimation Methods in Feedback Control Systems},
	year = {2014},
	publisher = {Wiley}
}

@article{WANORTBOB,
	author = {L. Wang and R. Ortega and A. Bobtsov},
	title = {Observability is sufficient for the design of globally exponentially convergent state observers for state-affine nonlinear systems},
	year = {2023},
	journal = {Automatica},
	volume = {149},
	pages = {110838},
}

%
\appendix
%
\section{Proof of Proposition \ref{pro2}}
\lab{appa}
%
Let $A\in\R^{n\times n}$ and $C \in \rea^n$ be in the observability canonical form
\[
A = \begin{bmatrix}
	-a_1 & 1 & 0 & \cdots & 0 \\
	-a_2 & 0 & 1 & \cdots & 0 \\
	\vdots & \vdots & \vdots & \ddots & \vdots \\
	-a_{n-1} & 0 & 0 & \cdots & 1 \\
	-a_n & 0 & 0 & \cdots & 0
\end{bmatrix},
\quad
C = \begin{bmatrix} 1 \\ 0 \\ \vdots \\ 0 \end{bmatrix}.
\]

It is clear from \eqref{dotphi} that $\Phi(t) = e^{At}$. Define the characteristic polynomial of $A$ as
\[
p(s)=s^n+a_1s^{n-1}+\cdots+a_n=\prod_{k=1}^n(s-\mu_k),\;\mu_k \in \mathbb{C}.
\]
We proceed now to evaluate the {\em impulse response} of the system. Towards this end, we need to consider the cases of {\em repeated} or {\em non repeated} eigenvalues. In the latter case we have
\[
p(s) = \prod_{m=1}^n (s-\mu_m)
\Longrightarrow
p'(\mu_k) = \prod_{\substack{m=1 \\ m\neq k}}^n (\mu_k - \mu_m) \neq 0,
\]
and the partial-fraction expansion is:
\[
\frac{1}{p(s)} = \sum_{k=1}^n \frac{1}{p'(\mu_k)}\cdot\frac{1}{s-\mu_k}.
\]
The associated impulse response of the system is given by
\[
z_0(t):=C^\top\Phi(t)=\sum_{k=1}^n \frac{1}{p'(\mu_k)}\,\e{\mu_k t}.
\]

On the other hand, when there are repeated eigenvalues, that is, when
\[
p(s) = \prod_{k=1}^r (s-\mu_k)^{m_k},
\qquad
m_1+m_2+\cdots+m_r = n.
\]
where the distinct eigenvalues be $\mu_1,\dots,\mu_r$ have multiplicities $m_1,\dots,m_r$. Then, the partial-fraction expansion becomes:
\[
\frac{1}{p(s)} = \sum_{k=1}^r \left[
\frac{A_{k,1}}{s-\mu_k} +
\frac{A_{k,2}}{(s-\mu_k)^2} + \cdots +
\frac{A_{k,m_k}}{(s-\mu_k)^{m_k}}
\right],
\]
where:
\[
A_{k,\nu} = \frac{1}{(m_k-\nu)!}\;\frac{d^{\,m_k-\nu}}{ds^{\,m_k-\nu}}\left[\frac{(s-\mu_k)^{m_k}}{p(s)}\right]_{s=\mu_k}.
\]
The impulse response for repeated eigenvalues is
\begin{align*}
	z_0(t)&=C^\top e^{A t} = \sum_{k=1}^r e^{\mu_k t} \Bigl[ A_{k,1}+A_{k,2}t+ \\
	&+A_{k,3}\frac{t^2}{2!}+ \cdots + A_{k,m_k}\frac{t^{m_k-1}}{(m_k-1)!}\Bigr].
\end{align*}
where we used 
$$
\mathcal{L}^{-1}\left\{\frac{1}{(s-\mu_k)^\nu}\right\} = \frac{t^{\,\nu-1}}{(\nu-1)!}\,e^{\mu_k t}.
$$

Now, the filtered signals satisfy the linear ODE:
\[
\dot{\Phi}_i^F = -\lambda_i \Phi_i^F + \lambda_i e^{A t}, \quad \Phi_i^F(0)={\bf 0}_{n \times n}.
\]
Integrating, yields
\[
\Phi_i^F(t) = \lambda_i e^{-\lambda_i t} \int_0^t e^{(\lambda_i I + A)\tau} d\tau,
\]
so:
\[
z_i(t) := C^\top \Phi_i^F(t) = \lambda_i e^{-\lambda_i t} \int_0^t z_0(\tau) e^{\lambda_i \tau} d\tau.
\]
Substituting $z_0(t) = \sum_{k=1}^n c_k e^{\mu_k t}$ we get
\[
z_i(t) = \sum_{k=1}^n \frac{\lambda_i c_k}{\lambda_i + \mu_k} \left(e^{\mu_k t} - e^{-\lambda_i t}\right),
\]
where, to simplify the notation, we defined 
$$
c_k: = {1 \over p'(\mu_k)} \neq 0.
$$
The matrix $Q(t)$ is then given as
\[
Q(t) = -\begin{bmatrix} z_0(t)  \\ z_1(t) \\ \vdots \\ z_{n-1}(t) \end{bmatrix},
\]
so $\det\{Q(t_k)\}\neq0$ if and only if the rows $\{z_0(t_k), \allowbreak z_1(t_k),\dots,z_{n-1}(t_k)\}$ are {\em linearly independent}.\\

We consider three different cases:
\begenu
\item[{\bf C1}] Eigenvalues of $A$ are {\em real and distinct}.
\item[{\bf C2}] Eigenvalues of $A$ are {\em real but not distinct}.
\item[{\bf C3}] There are {\em complex} eigenvalues.
\endenu

\subsection*{Case 1}

Rewrite all signals as:
\begalis{
	z_0(t) &= \sum_{k=1}^n c_k e^{\mu_k t} \\
	z_i(t) &= \sum_{k=1}^n \frac{\lambda_i c_k}{\lambda_i + \mu_k} e^{\mu_k t} - \left(\sum_{k=1}^n \frac{\lambda_i c_k}{\lambda_i + \mu_k}\right) e^{-\lambda_i t}.
}

The set of exponential functions $\{e^{\mu_1 t},\dots,e^{\mu_n t},e^{-\lambda_1 t},\allowbreak \dots,e^{-\lambda_{n-1} t}\}$ is linearly independent over $(0,\infty)$ if and only if all exponents are distinct. Linear {\em dependence} occurs if and only if some $\lambda_i = -\mu_k$, a case that is ruled out under the standing assumptions. On the other hand, because of the assumption of stability of $A$, we see that we have $\det\{Q(t)\}=0$ at $t=\infty$.

\subsection*{Case 2}

Let the distinct eigenvalues be $\{\mu_1,\dots,\mu_r\}$ with multiplicities summing to $n$. Repeated eigenvalues produce polynomial--exponential terms:
\[
e^{\mu t},\ t e^{\mu t},\ t^2 e^{\mu t},\ \dots,\ t^{m-1} e^{\mu t}.
\]
The impulse response becomes:
\[
z_0(t) = \sum_{k=1}^r q_k(t) e^{\mu_k t},
\]
where $q_k(t)$ is a polynomial of degree $\le m_k-1$. The family of functions $\{t^s e^{\mu_k t}\}\cup\{e^{-\lambda_i t}\}$ is linearly independent if and only if all exponents are distinct and no $\lambda_i = -\mu_k$ that, again, is ruled out. We conclude then that repeated eigenvalues do not destroy independence.

\subsection*{Case 3}

Let a conjugate pair of non-real eigenvalues be:
\[
\mu = \alpha + j\beta,\quad \overline{\mu} = \alpha - j\beta,\quad \alpha,\beta\in\rea,\ \beta>0.
\]
The condition for linear independence of the rows of $Q(t)$ given above is  $\lambda_i+\mu=0$, which is impossible to hold because $\beta>0$. Consequently $\detQ \equiv 0$ {\em cannot} happen. However, we will show below that $\detQ$ {\em may be} equal to zero for some {\em isolated times} $t_j>0$. 

The terms corresponding to $\mu$ and $\overline{\mu}$ in the determinant sum are:
\[
\frac{\e{(\alpha+j\beta)t}}{\prod_{i=1}^{n-1}(\lambda_i+\alpha+j\beta)}
+
\frac{\e{(\alpha-j\beta)t}}{\prod_{i=1}^{n-1}(\lambda_i+\alpha-j\beta)}.
\]
For simplicity, take $n=2$ (one $\lambda_1$):
\begin{align*}
&\frac{\e{(\alpha+j\beta)t}}{\lambda_1+\alpha+j\beta}
+
\frac{\e{(\alpha-j\beta)t}}{\lambda_1+\alpha-j\beta}
= \\
&=\frac{2\,\e{\alpha t}}{(\lambda_1+\alpha)^2+\beta^2}
\bigl[(\lambda_1+\alpha)\cos\beta t + \beta\sin\beta t\bigr].
\end{align*}
We consider, again, two cases. First, when $\lambda_1 \neq -\alpha$. In this case, both coefficients $(\lambda_1+\alpha)$ and $\beta$ are non-zero.
The trigonometric combination:
\[
(\lambda_1+\alpha)\cos\beta t + \beta\sin\beta t
\]
is a \emph{sinusoidal} signal---it vanishes at \emph{isolated times} but {never identically} for all $t\ge0$. 

The second case is when $\lambda_1 = -\alpha$. Then, the expression simplifies to $\frac{2\,\e{-\lambda_1 t}}{\beta}\,\sin\beta t$, 
This term {\em vanishes} at $t = {k\pi \over \beta}$, $k=0,1,2,\dots$, but does not vanish for all $t\ge0$. 

This completes the proof.
%
\section{Proof of Lemma \ref{lem1}}
\lab{appb}
%
\subsection{Evaluation of $\Phi(t)$ for the non-holonomic system \eqref{nonholsysA}-\eqref{nonholsysB}, \eqref{u1u2}}
Consider the matrix ODE:
\[
\dot{\Phi}(t) = A(t)\,\Phi(t), \quad \Phi(0)={\bf I}_4,
\]
where
\[
A(t) = \block{0_{2\times2}}{A_{12}(t)}{0_{2\times2}}{A_{22}(t)},
\quad
A_{12}(t)=e^{-t}{\bf I}_2,
\]
\[
A_{22}(t)=\begin{bmatrix}0 & -f(t) \\ f(t) & 0\end{bmatrix},
\] 

and
\[
f(t)=2e^{-t}+2te^{-t}.
\]

Define the phase function:
\[
g(t)=\int_0^t f(\tau)\,d\tau = 4 - 2(t+2)e^{-t}.
\]
Then the solution is block-triangular:
\[
\Phi(t) = \bigblock{{\bf I}_2}{\Phi_{12}(t)}{0_{2\times2}}{R(g(t))},
\]
where the rotation matrix is
\[
R(g)=\begin{bmatrix}\cos g & -\sin g \\ \sin g & \cos g\end{bmatrix},
\]
and the off-diagonal block satisfies
\[
\Phi_{12}(t)=\begin{bmatrix}C(t) & -S(t) \\ S(t) & C(t)\end{bmatrix}
\]
with
\[
C(t)=\int_0^t e^{-\tau}\cos g(\tau)\,d\tau, \quad
S(t)=\int_0^t e^{-\tau}\sin g(\tau)\,d\tau.
\]
These integrals do not admit elementary closed form and are evaluated numerically.

\subsection{Evaluation of $\Phi^F(t)$}
Solve
\[
\dot{\Phi}^F(t) = -\lambda\Phi^F(t)+\lambda\Phi(t),\; \Phi^F(0)={\bf 0}_{4 \times 4},\;\lambda>0.
\]

Using the integrating factor $e^{\lambda t}$, the solution is:
\[
\Phi^F(t)=\lambda\,e^{-\lambda t}\int_0^t e^{\lambda\tau}\,\Phi(\tau)\,d\tau.
\]
By block structure:
\[
\Phi^F(t)=\block{\Phi^F_{11}(t)}{\Phi^F_{12}(t)}{\Phi^F_{21}(t)}{\Phi^F_{22}(t)},
\]
where
\[
\Phi^F_{11}(t)=(1-e^{-\lambda t})I_2,\qquad \Phi^F_{21}(t)\equiv 0,
\]
\[
\Phi^F_{22}(t)=\lambda \,e^{-\lambda t}\begin{bmatrix}U_c(t) & -U_s(t) \\ U_s(t) & U_c(t)\end{bmatrix},
\]
\[
\begin{aligned}
	U_c(t)&=\int_0^t e^{\lambda \tau}\cos g(\tau)\,d\tau,\\
	U_s(t)&=\int_0^t e^{\lambda \tau}\sin g(\tau)\,d\tau,
\end{aligned}
\]
and
\[
\Phi^F_{12}(t)=\begin{bmatrix}C(t)-e^{-\lambda t}K_c(t) & -S(t)+e^{-\lambda t}K_s(t) \\[4pt]
	S(t)-e^{-\lambda t}K_s(t) & C(t)-e^{-\lambda t}K_c(t)\end{bmatrix},
\]
with
\[
K_c(t)=\int_0^t e^{(\lambda -1)\tau}\cos g(\tau)\,d\tau,
\]
\[
K_s(t)=\int_0^t e^{(\lambda -1)\tau}\sin g(\tau)\,d\tau.
\]

\subsection{Evaluation of the rank of $Q(t)$}
Define
\[
Q(t)=\block{I_2}{\Phi_{12}(t)}{\Phi^F_{11}(t)}{\Phi^F_{12}(t)}.
\]

Using the block-determinant formula:
\begin{align*}
\det Q(t)&=\det\!\left(\Phi^F_{12}(t)-\Phi^F_{11}(t)\Phi_{12}(t)\right) =\\
&= \det\!\left(-e^{-\lambda t}K(t)\right),
\end{align*}
where
\[
K(t)=\begin{bmatrix}K_c(t) & -K_s(t) \\ K_s(t) & K_c(t)\end{bmatrix}.
\]
Thus
\[
{\det Q(t)=e^{-2\lambda t}\left(K_c(t)^2+K_s(t)^2\right)}.
\]

We have the following conclusion.
\begin{itemize}
	\item At $t=0$: $K_c(0)=K_s(0)=0 \implies \det Q(0)=0$, so $\rank Q(0)=2$ (singular).
	\item For all $t>0$: $K_c(t)^2+K_s(t)^2>0$ almost everywhere, hence $\rank Q(t)=4$ (full rank).
	\item As $t\to\infty$, $\det Q(t)\to 0$.
\end{itemize}
In conclusion $Q(t)$ is singular only at $t=\{0,\infty\}$ and is full rank for all $\infty>t>0$.

\end{document}